\documentclass[pra, notitlepage, superscriptaddress,  reprint]{revtex4-1}
\usepackage{comment}
\usepackage{enumerate}
\usepackage{amssymb}
\usepackage{amsmath}
\usepackage{braket}
\usepackage{graphicx}
\usepackage[usenames,dvipsnames]{color}
\usepackage{tensor}
\usepackage{empheq}
\usepackage{capt-of}
\usepackage[normalem]{ulem} 
\usepackage{soul} %

\usepackage[colorlinks,bookmarks=false,citecolor=NavyBlue,linkcolor=OliveGreen,urlcolor=blue]{hyperref}
\newcommand{\be}{\begin{equation}}
\newcommand{\ee}{\end{equation}}
\newcommand{\ba}{\begin{aligned}}
\newcommand{\ea}{\end{aligned}}
\newcommand{\bw}{\begin{widetext}}
\newcommand{\ew}{\end{widetext}}

\newcommand{\bea}{\begin{eqnarray}}
\newcommand{\eea}{\end{eqnarray}}

\newcommand{\tr}[2]{\mathrm{tr}_{#1}(#2)}

\def\doi{http://dx.doi.org/}

\begin{document}
\title{Transport in out-of-equilibrium XXZ chains: \\
non-ballistic behavior and correlation functions}
\author{Lorenzo Piroli}
\address{SISSA and INFN, via Bonomea 265, 34136, Trieste, Italy}
\author{Jacopo De Nardis}
\address{D\'epartement de Physique, \'Ecole Normale Sup\'erieure / PSL Research University, CNRS, 24 rue Lhomond, 75005 Paris, France}
\author{Mario Collura}
\address{The Rudolf Peierls Centre for Theoretical Physics, Oxford University, Oxford, OX1 3NP, United Kingdom}
\author{Bruno Bertini}
\address{SISSA and INFN, via Bonomea 265, 34136, Trieste, Italy}
\author{Maurizio Fagotti}
\address{D\'epartement de Physique, \'Ecole Normale Sup\'erieure / PSL Research University, CNRS, 24 rue Lhomond, 75005 Paris, France}

\begin{abstract}
We consider the nonequilibrium protocol where two semi-infinite gapped 
XXZ chains, initially prepared in different equilibrium states, are suddenly joined together. At large times, a generalized hydrodynamic description applies,  according to which the system can locally be represented by space- and time- dependent stationary states. The magnetization displays an unusual behavior: depending on the initial state, its profile may exhibit abrupt jumps that can not be predicted directly from the standard hydrodynamic equations and which signal non-ballistic spin transport. We ascribe this phenomenon to the structure of the local conservation laws and make a prediction for the exact location of the jumps. We find that the jumps propagate at the velocities of the heaviest quasiparticles.  By means of time-dependent density matrix renormalization group simulations we show that our theory yields a complete description of the long-time steady profiles of conserved charges, currents, and local correlations.
\end{abstract}

\maketitle

The role of integrability in modern many-body quantum physics could hardly be overestimated, having strengthened our understanding of ground-state and thermal physics for the better part of the last century \cite{books,takahashi,gaudin,korepin}. In the past two decades there has been an unprecedented and increasing interest in the nonequilibrium dynamics of isolated systems, and integrable models have been the ideal environment where to investigate out-of-equilibrium physics (see \cite{CaEM16} for a volume of recent pedagogic reviews on the subject). This is  intimately connected with  the development of new experimental techniques in cold atoms, which now allow us to probe almost unitary quantum nonequilibrium dynamics~\cite{bdz-08,ccgo-11,pssv-11,LaGS16,kww-06,HLFS07,gklk-12,fse-13,lgkr-13,glms-14,langen-15}. 

One problem that has catalyzed an impressive amount of theoretical efforts is the characterization of the long-time steady state reached in an integrable system prepared in an out-of-equilibrium state. In the simplest instance, the system is initially prepared in a homogeneous equilibrium state, and perturbed by a sudden change in one of the Hamiltonian parameters (a so-called quantum quench \cite{cc-06}). It is now  established that local observables relax and can be quantitatively described by a Generalized Gibbs Ensemble (GGE), a statistical ensemble built taking into account all the local and quasi-local conserved operators~\cite{rdyo-07,ViRi16,EsFa16}.

\begin{figure}[h]

\includegraphics[width=0.49\textwidth]{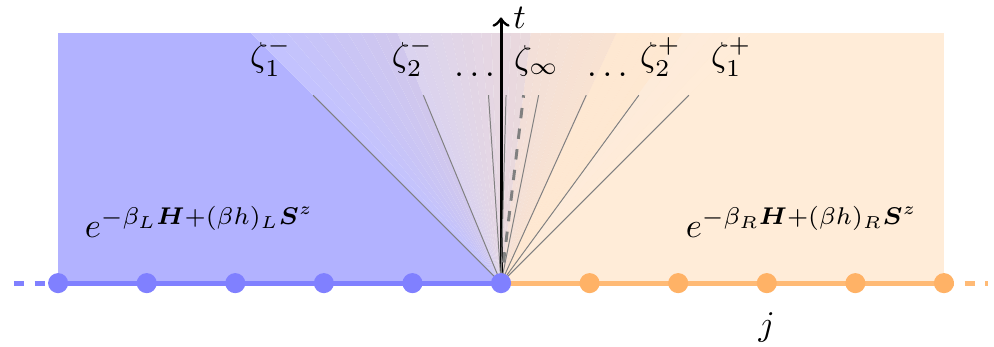}\vspace{-0.25cm}
\caption{Pictorial representation of the quench protocol. After joining together two semi-infinite XXZ chains, quasiparticle excitations are created. Different types of quasiparticles move with different maximum and minimum velocities $\zeta^{\pm}_n$. The heaviest quasiparticles move with velocity $\zeta_\infty$ (\emph{cf.} Sec.~\ref{sec:lightcones}).  
 }
\label{fig:lightcone_cartoon}
\end{figure}

More recently, increasing attention has been devoted to the more general and complex case where the system is initially prepared in a inhomogeneous state, for example by joining together two macroscopically different homogeneous states. Initially, analytical understanding was restricted to either free systems \cite{ARRS99,AsPi03,AsBa06,PlKa07,LaMi10,EiRz13,DVBD13,Bert17,ADSV16,
CoKa14,EiZi14,CoMa14,DeMV15,DLSB15,VSDH16,KoZi16,Korm17,PeGa17} or conformally invariant models and Luttinger liquids \cite{SoCa08,CaHD08,Mint11,MiSo13,DoHB14,BeDo12,BeDo15,BeDo16,BeDo16Review,LLMM17,DuSC17, SDCV17}, while  genuinely interacting systems were mainly investigated numerically, or relying on \emph{ad hoc} conjectures \cite{GKSS05,SaMi13,DVMR14,AlHe14,CCDH14,BDVR16,Zoto16,ViIR17}. A breakthrough came in the past year, with the introduction of the so-called generalized hydrodynamics \cite{CaDY16,BCDF16}. A significant number of studies have already been devoted to further investigate some of its most interesting aspects \cite{DoYo16,DoSY17,DoSp17,DoYC17,DDKY17,BVKM17-2,BVKM17, F16}. Indeed,
such approach is extremely flexible, allowing one to study, for instance, nonequilibrium dynamics in the presence of localized defects \cite{BeFa16,DeLucaAlvise}, or of confining traps \cite{DoYo16}. Furthermore, it has been shown that the hydrodynamic picture not only gives an exact description at infinite length- and time- scales but can also be a surprisingly good approximation  at finite scales~\cite{DDKY17,BVKM17,BVKM17-2} or in the presence of small integrability breaking terms~\cite{DoYo16}. These developments also boosted the study of linear and non-linear transport in integrable systems,  which is a topic of long-standing interest and with a long history ~\cite{CaZP95,HHCB03,SiPA09,Pros11,Znid11,StBr11,KaBM12,PrIl13,LHMH11,PrZn09,KaMH14,
VaKM15,Doyo15,SHZB16,Karr17,PBPD17}. Hydrodynamic approaches led to important results on several open questions, such as the nature of spin and charge Drude weights~\cite{IlDe17,DoSp17-2}. In addition, related ideas were recently employed for the computation of the time evolution of the entanglement entropy after a global quench~\cite{AlCa16}.

According to the hydrodynamic picture, at large times $t$ local subsystems at distance $x$ from the junction reach different stationary states depending on the ``ray" $x/t$, see Fig.~\ref{fig:lightcone_cartoon}. Stationary states describing observables on a fixed ray are characterized by appropriate GGEs or, equivalently, by appropriate distributions of quasi-momenta (or \emph{rapidities}) of the elementary excitations. 

The derivation of the hydrodynamic equations proposed in \cite{CaDY16} and \cite{BCDF16} is based on the existence of a complete set of conservation laws. 
Remarkably, the conservation laws in the XXZ chain have a different structure in the gapped and gapless regimes \cite{PrIl13,prosenreview,IDWC15,IlievskiJSTAT,PiVC16,ZaMP16,DeLuca16}. 
Only the latter case, however, has been analyzed theoretically~\cite{BCDF16}, so it is natural to wonder whether and why qualitative differences emerge in the space-time profiles of local observables. 

In this work we settle this issue. We provide a detailed analysis of the space-time profiles of local conserved charges, currents and local correlations in the gapped regime. The most remarkable phenomena are observed when the sign of the magnetization in the initial state is not the same on the two sides of the junction. In that case, observables display a different behavior depending on how they transform under spin flip. In particular, the magnetization exhibits abrupt jumps, which can not be predicted directly from the hydrodynamic equations obtained in \cite{BCDF16}; the jumps are the signature of non-ballistic transport. We derive an equation that describes the location of the jumps, and relate them to the velocity of the heaviest quasiparticles. 
This information complements the structure unveiled in \cite{BCDF16},  so as to provide a complete description of the long-time steady profiles of all local observables in the gapped regime. 
Moreover, we discuss the emergence of non-analyticities in the profiles of observables, revealing their connection with the quasiparticle content of the theory. 

The manuscript is laid out as follows. In Section~\ref{sec:themodel} we introduce the XXZ model and its thermodynamic Bethe ansatz (TBA) solution. In Section~\ref{sec:protocol} we describe the quench protocol and review the hydrodynamic approach. Section~\ref{sec:main} shows how to determine the jumps in observables which are odd under spin flip. In Section~\ref{sec:results} we analyze profiles of densities of charges, currents and correlation functions, and discuss the implications of the jumps for the spin transport. Section~\ref{sec:conclusions} contains our conclusions.

\section{The model}\label{sec:themodel}%

We consider the XXZ spin-$1/2 $ chain described by the Hamiltonian
\bea
\boldsymbol H &=& \sum_{j=-\frac{L}{2}}^{\frac{L}{2}-1}\left[\boldsymbol{s}_{j}^{x}\boldsymbol {s}_{j+1}^{x}+\boldsymbol{s}_{j}^{y}\boldsymbol{s}_{j+1}^{y}+ \Delta  \boldsymbol{s}_{j}^{z}\boldsymbol{s}_{j+1}^{z}\right]\,,
\label{eq:Hamiltonian_XXZ}
\eea
where   $\boldsymbol{s}_j^\alpha =\frac{1}{2}\boldsymbol{\sigma}_j^\alpha $ ($\alpha=x,y,z$) and $\boldsymbol{\sigma}_j^\alpha$ are Pauli matrices. We assume periodic boundary conditions, namely ${\boldsymbol{s}_{\frac{L}{2}}^\alpha = \boldsymbol{s}_{-\frac{L}{2}}^\alpha}$. 
We focus on the case ${\Delta> 1}$, where the Hamiltonian \eqref{eq:Hamiltonian_XXZ} is gapped, and make use of the parametrization
\bea
\Delta=\cosh\eta\,,\qquad\qquad \eta >0 \,.\label{eq:def_eta}
\eea
We denote the energy density operator by $\boldsymbol e_j$, \emph{i.e.} ${\boldsymbol H=\sum_j \boldsymbol e_j}$.
The eigenvalues of \eqref{eq:Hamiltonian_XXZ} can be constructed exactly by means of the Bethe ansatz method~\cite{Bethe, takahashi}.   
In that framework, eigenstates are characterized in terms of rapidities $\lambda\in [-\pi/2,\pi/2]$ of the so-called magnons, fulfilling some appropriate quantization conditions. As the theory is fully interacting, these particles can create bound states of any size $n$~\cite{takahashi} ($n=1$ corresponds to unbound magnons). In the thermodynamic limit $L \to \infty$, the number of magnons diverges and the stationary states with nonnegative magnetization (we will come back to this point later) are characterized by the densities of their quasi momenta. 
For each bound state type there is a density of rapidities (or root density) $\rho_n(\lambda)$. These densities are such that, in a large finite volume $L$, $L\rho_n(\lambda){\rm d}\lambda$ gives the number of magnonic bound states of length $n$ with rapidities in the interval $[\lambda, \lambda + d\lambda)$. The root densities can be thought of as the generalization to interacting models of the occupation numbers in free systems. In the XXZ chain with $\Delta>1$, they characterize the expectation value of any local operator which is invariant under spin flip (see the discussion in Section~\ref{subsec:charges}). 

Along with the rapidity distributions $\rho_n(\lambda)$, it is customary to introduce the hole distribution functions $\rho_n^{h}(\lambda)$; they describe allowed values of the rapidities which are not occupied in the state. Distributions of particles and holes are related by the TBA equations 
\begin{equation}\label{eq:BetheTaka_gapped}
\rho_n(\lambda)+\rho_n^h(\lambda) =a_n(\lambda) - \sum_{m=1}^\infty \,(T_{nm} \ast \rho_m)(\lambda)\,.
\end{equation}
Here we introduced the so called scattering kernel $T_{nm} (\lambda)$, which reads as 
\bea
T_{nm}(\lambda)&=&(1-\delta_{nm})a_{|n-m|}(\lambda)+2a_{|n-m|+2}(\lambda)\nonumber\\
&+&\ldots +2a_{n+m-2}(\lambda)+a_{n+m}(\lambda)\,,
\eea
where the functions $a_n(\lambda)$ are given by 
\be
a_n(\lambda)=\frac{1}{\pi} \frac{\sinh\left( n\eta\right)}{\cosh (n \eta) - \cos( 2 \lambda)}\,.
\label{def:a_function_gapped}
\ee
The convolution between two functions is defined as follows
\be
\left(f\ast g\right)(\lambda)=\int_{-\pi/2}^{\pi/2}{\rm d}\mu f(\lambda-\mu)g(\mu)\,.
\label{eq:convolution}
\ee
Given the distributions $\rho_n(\lambda)$ and $\rho_n^h(\lambda)$, the following combinations
\be
\rho_n^t(\lambda)=\rho_n(\lambda)+\rho_n^h(\lambda)\,,\quad
\vartheta_n(\lambda)=\frac{\rho_n(\lambda)}{\rho^{t}_n(\lambda)}\,,\label{eq:theta_function}
\ee
are usually called total root densities and filling functions, respectively. 

\subsection{Conserved quantities and rapidity distributions}\label{subsec:charges}%

Being integrable, the model described by the Hamiltonian~\eqref{eq:Hamiltonian_XXZ} admits a macroscopically large set of local and quasi-local conserved charges $\{\boldsymbol S^z, \boldsymbol Q^{(n)}_{j}\}$~\cite{prosenreview}.
Here $\boldsymbol S^z=\sum_{\ell}\boldsymbol s_\ell^z$ indicates the total magnetization in the $z$ direction and  $\boldsymbol Q_{1}^{(1)}+\frac{\Delta L}{4}$ is the Hamiltonian \eqref{eq:Hamiltonian_XXZ}. The expectation values of these charges can be taken as the quantum numbers used to  classify the eigenstates of the Hamiltonian in the thermodynamic limit. As shown in \cite{IDWC15,IlievskiJSTAT}, there is a one-to-one correspondence between the distribution of rapidities $ \rho_n(\lambda)$ and the conserved quantities. In the gapped regime, given the expectation values of the charges, the distributions $\rho_n(\lambda)$ read as 
\begin{equation}
\rho_n(\lambda) = X_n^{+}(\lambda) + X_n^{-}(\lambda) - X_{n-1}(\lambda) - X_{n+1}(\lambda)\,.
\label{eq:chargesrho}
\end{equation} 
Here ${n>0}$,  the quantities $X_n(\lambda)$ are the generating functions of the expectation values of $\{\boldsymbol Q^{(n)}_j\}$, and ${X_n^{[\pm]}(\lambda)=X_n(\lambda \pm i \eta/2)}$. Importantly, equation \eqref{eq:chargesrho} can be ``inverted": given the distributions $\{ \rho_n \}$, the expectation values of the density $\boldsymbol q$ of a generic conserved charge $\boldsymbol Q \in \{\boldsymbol Q_j^{(n)}\}$ is given by  
\be\label{eq:q}
\langle \boldsymbol{q} \rangle =\sum_{n=1}^\infty \int\mathrm d\lambda \ q_n(\lambda)\rho_n(\lambda)\,,
\ee
where the functions $q_n(\lambda)$ are called single-particle eigenvalues or bare charges. For example, the single-particle eigenvalues of the energy density (shifted by ${\Delta}/{4}$) are  $q_n(\lambda)  = - \pi \sinh(\eta) a_n(\lambda)$.

For ${\Delta>1}$, the conserved charges generated by $X_n(\lambda)$ are invariant under spin flip $\boldsymbol{\mathcal O}\rightarrow \Pi \boldsymbol{\mathcal O}\Pi $, with ${\Pi = \prod_{i} \boldsymbol{\sigma}^x_i}$, so the  functions $X_n(\lambda)$  do not change if they are computed in the state where all the spins are flipped. As a result, the stationary states with magnetization $\langle \boldsymbol{S}^z \rangle$ and  $-\langle \boldsymbol{S}^z \rangle$ are described by the same set of rapidity distributions $\rho_n(\lambda)$ \cite{IlievskiJSTAT,PiVC16}. This is a crucial difference with respect to  the regime $|\Delta|<1$, where also odd conserved charges are  generated by some $X_n(\lambda)$, and states with opposite magnetization are described by different distributions $\rho_n(\lambda)$ \cite{DeLuca16,IlDe17}.
For $\Delta>1$, $\rho_n(\lambda)$ are sufficient to characterize only the expectation values of the observables which are even under spin flip, including the absolute value of the magnetization
\begin{align}
\label{eq:magn-rhos} 
|\langle \boldsymbol{s^z} \rangle| &= \frac{1}{2} - \sum_{n=1}^\infty \int\mathrm d\lambda \ n \ \rho_n(\lambda)\notag\\
&= \lim_{n \to \infty} \frac{1}{2}\int d\lambda \ \rho_n^h(\lambda)\geq 0\,,
\end{align}
where in the second step we used the TBA equations \eqref{eq:BetheTaka_gapped}. 

Since, in the present understanding,  $\boldsymbol S^z$ is the only odd conserved charge, only an additional ``bit" of information is required to fully characterize the state. Specifically, it is widely accepted that it is sufficient to supplement the set of $\rho_n(\lambda)$ with a binary variable $\mathfrak{f}=\pm$, which bears information about the sign of the magnetization. We indicate by $\ket{\rho,\mathfrak{f}}$ a state with sign of the magnetization equal to $\mathfrak{f}$ and rapidity distributions given by $\rho_n(\lambda)$. Expectation values in the state $\ket{\rho,\mathfrak{f}}$ are denoted by $\braket{\cdot}^{\frak f}$. For operators $\boldsymbol{\cal O}_{\rm e}$, even under spin flip, we have 
\be
\braket{\boldsymbol{\cal O}_{\rm e}}^{\frak f} = \braket{\boldsymbol{\cal O}_{\rm e}}^+\,,
\ee
while for odd operators $\boldsymbol{\cal O}_{\rm o}$ we have   
\be
\braket{\boldsymbol{\cal O}_{\rm o}}^{\frak f} = \frak f \braket{\boldsymbol{\cal O}_{\rm o}}^+\,.
\ee       
Note that $\braket{\boldsymbol{s^z}}^{+}$ is that given in Eq.~\eqref{eq:magn-rhos}.

\subsection{Universal dressing equations and velocities}%

Let us consider the system in a large, finite volume $L$. In the thermodynamic limit, the state is described by the root densities $\rho_n(\lambda)$, or, equivalently, by the filling functions $\vartheta_n(\lambda)$. Excitations on this state can be constructed by injecting an extra string of size $n$ with rapidity $\lambda$. This operation induces a change in the expectation values of the conserved charges  
\begin{align}
& \langle \boldsymbol{Q}\rangle^{\frak f} \to \langle \boldsymbol{Q} \rangle^{\frak f} + q_n^d(\lambda)\, .
\end{align}
Here $q_n^d(\lambda)$ is called  ``dressed charge" and is an $O(L^0)$ deformation of the charge due to the presence of the new particle of type $n$ with rapidity $\lambda$. Its derivative with respect to $\lambda$, $q^{d\,\prime}_n(\lambda)$, can be expressed as a linear integral equation which takes the following universal form
\be
q_n^{\prime\,d} (\lambda)=q_n^{\prime}  (\lambda) - \Big[ \sum_{m=1}^\infty \,T_{nm} \ast ( q_m^{\prime\,d}  \vartheta_m) \Big]  (\lambda)\,.
\label{eq:dressingINTRO}
\ee
Here ${q_n(\lambda)}$ is the bare charge (\emph{cf}. Eq~\eqref{eq:q}), \emph{i.e.}, the charge computed with respect to the reference state with all spins up. We note that the bare charge $p_n(\lambda)$ for the momentum is such that $p_n^{\prime}  (\lambda)=2\pi a_n(\lambda)$, so from Eq.~\eqref{eq:dressingINTRO} and Eq.~\eqref{eq:BetheTaka_gapped} it follows
\be
p_n^{\prime\,d} (\lambda)= 2 \pi \rho_n^t(\lambda)\, .
\ee
We indicate with  $\varepsilon_n(\lambda)$ the dressed energy of the particle excitations.
From the momentum and the energy we can extract the group velocity of a particle excitation of type $n$ and rapidity $\lambda$ \cite{bonnes14}
\begin{equation}\label{eq:velocityINTRO}
v_n (\lambda) = \frac{\partial \varepsilon_n(\lambda) }{\partial p_n^d(\lambda)} = \frac{\varepsilon_n^{\prime} (\lambda)}{2 \pi  \rho_n^t (\lambda)} .
\end{equation}
We stress that the dressing equations \eqref{eq:dressingINTRO} are valid for any integrable model (with diagonal scattering), provided that its scattering kernel $T_{nm}(\lambda)$ is known. 

\subsection{Currents} %

A current $\boldsymbol{J}_{q,\ell}$ is defined in terms of the density of charge $\boldsymbol q_\ell$ via the following continuity equation 
\be
\boldsymbol{J}_{q,\ell+1}-\boldsymbol{J}_{q,\ell}=i[\boldsymbol q_\ell,\boldsymbol H]\,.
\ee
Requiring $\boldsymbol{J}_{q,\ell}$ to vanish in the reference state, the operator $\boldsymbol{J}_{q,\ell}$ is determined up to operators with zero expectation value in any translationally invariant state. Importantly, currents are generically \emph{not} conserved and, after a quantum quench, their expectation values undergo a non trivial time evolution. 

An important result of Refs~\cite{BCDF16,CaDY16} was to suggest how to compute the expectation value of a current in a ``generic'' stationary state in generic TBA solvable systems. The result takes the simple form 
\be\label{eq:j}
\langle \boldsymbol{J}_q \rangle=\sum_{n=1}^\infty  \int\mathrm d\lambda \ q_n(\lambda)v_n(\lambda) \rho_n(\lambda)\, .
\ee 
For $\Delta>1$, this expression applies to the current of every charge but $\boldsymbol S^z$. In particular, in the case of the energy current this expression can be rewritten as \cite{BCDF16}
\begin{align}
\langle \boldsymbol{J}_e \rangle^{\frak f}& =  \sum_{n=1}^\infty \int\!\!\mathrm d\lambda \ e_n(\lambda)\ \rho_n(\lambda) v_n(\lambda)\notag\\
&= \sum_{n=1}^\infty \int\!\!\mathrm d\lambda \ q^{(1)}_{n,2}(\lambda)\ \rho_n(\lambda) = \langle \boldsymbol{q}_2^{(1)} \rangle^{\frak f}\,,
 \label{eq:Jen-rhos}
\end{align}
where $q^{(1)}_{2,n}(\lambda)$ is the bare charge corresponding to the second local conserved charge $\boldsymbol Q_2^{(1)}$. This is in accordance with the well-known relation $\sum_\ell \boldsymbol{J}_{e,\ell} = \boldsymbol Q_2^{(1)}$. The spin current has to be supplemented with the information on the sign of the magnetization (\emph{cf}. Sec.~\ref{subsec:charges}) and reads as
\begin{align}
\langle \boldsymbol{J}_s \rangle^{\frak f}  & = \frak f \sum_{n=1}^\infty \int\mathrm d\lambda \ n\ \rho_n(\lambda) v_n(\lambda)\nonumber \notag\\
& = \frac{\frak f }{2} \lim_{n \to \infty}\int d\lambda \rho^h_n(\lambda) v_n(\lambda)\,;\label{eq:Jspin-rhos}
\end{align}
in the second step, we used the equations \eqref{eq:dressingINTRO}.

\section{The quench protocol and the hydrodynamic equations}\label{sec:protocol}%

\subsection{The Initial State}  %

As discussed in the introduction, in this work we consider the nonequilibrium dynamics resulting from joining two semi-infinite chains with different macroscopic properties. In particular, we focus on the case where two chains are at thermal equilibrium with different values of temperature and magnetic field. The initial state is then given by
\begin{equation}\label{eq:ini_state}
\boldsymbol \rho_0 = \frac{e^{- \beta_L \boldsymbol{H}_L + (\beta h)_L \boldsymbol{S}_L^z }}{Z_L} \bigotimes \frac{e^{- \beta_R \boldsymbol{H}_R+ (\beta h)_R \boldsymbol{S}_R^z }}{Z_R}\,,
\end{equation}
where the operators with the subscript $L$ ($R$) are defined by restricting the sums of their density to the negative (positive) sites, while $Z_L$ and $Z_R$ are appropriate constants that ensure normalization. 

Starting from $\boldsymbol \rho_0$, the region where local observables are thermal remains macroscopically large: at any time $t$,  as a consequence of the Lieb-Robinson bounds~\cite{LR72}, far away from the junction local observables are always described by thermal states. In integrable models, however, there is a region of width~${\sim t}$ around the origin where observables are described by a family of non-thermal stationary states, as pictorially represented in Fig.~\ref{fig:lightcone_cartoon}. The characterization of this family, which was called Locally Quasi-Stationary State in \cite{BeFa16}, is the subject of the next subsection.

\subsection{The Hydrodynamic Equations}\label{sec:hydrodynamiceq}%

In integrable models, like the XXZ spin-1/2 chain, there are stable quasiparticle excitations which propagate at different velocities and scatter elastically with one another. They are responsible for the propagation of information throughout the system~\cite{bonnes14}. In many cases of interest, at large time- and length- scales, the quasiparticle excitations behave like free classical particles, and the effects of the interactions can be taken into account by letting the velocity of the quasiparticles  to depend on the state. 
In particular, if the initial state is the junction of two homogeneous states, one can infer that, at large times, local observables moving on a certain ``ray" $\zeta=j/t$ are characterized by a $\zeta$-dependent steady state $\boldsymbol \rho_s (\zeta)$, \emph{c.f.} Fig.~\ref{fig:lightcone_cartoon}. Indeed, different rays receive information from different quasiparticles. The hydrodynamic equations are derived under this assumption. 

We note that the state becomes equivalent to $\boldsymbol \rho_s (\zeta)$  only when both the time and the distance from the junction approach infinity at fixed ratio. By fixing the position and increasing time, the observables explore the entire family of stationary states, eventually ending up in the state $\boldsymbol \rho_s (0)$. The state $\boldsymbol \rho_s (0)$ is known as nonequilibrium steady state (NESS). 

The state $\boldsymbol \rho_s (\zeta)$ has been characterized in Refs.~\cite{BCDF16,CaDY16}, using that the expectation values of the local and quasi-local charges determine the rapidity distributions. Specifically, the root densities $\rho_{n, \zeta}(\lambda)$ of the state  $\boldsymbol \rho_s (\zeta)$ have been shown to satisfy the following continuity equation
\be\label{eq:cont}
\zeta \partial_{\zeta} \rho_{n,\zeta}(\lambda)=\partial_\zeta\Bigl[v_{n,\zeta}(\lambda) \rho_{n,\zeta}(\lambda)\Bigr]\,,
\ee
where the velocity $v_{n,\zeta}(\lambda)$ is given in Eq.~\eqref{eq:velocityINTRO}. Here we are working in the limit of infinite times and distances at fixed ray $\zeta$, where Eq.~\eqref{eq:cont} is exact. One could also try to extend this equation to describe finite-time dynamics. However, further terms would appear. In particular, we can easily identify two kinds of finite-time corrections to the naive finite-time version of \eqref{eq:cont}
\be\label{eq:contxt}
\partial_t\rho_{n,x,t}+\partial_x \Bigl[v_{n}[\rho_{n,x,t}]\rho_{n,x,t}\Bigr]=0\, .
\ee
The first type of corrections is  related to the introduction of finite length scales, which make 
the thermodynamic description only approximate. While such corrections could in principle be written in terms of root densities, it is difficult to estimate them in practice. The second kind of corrections are due to the fact that currents are not generically conserved. Indeed, as discussed in \cite{F16} for the specific case of a noninteracting model, the expectation values of the currents  take the form \eqref{eq:j} only if the state is stationary. These corrective terms can not be generically written in terms of root densities. Nevertheless, for particular classes of initial states, such corrections might be very small, leading to accurate quantitative predictions~\cite{BVKM17-2,BVKM17,DDKY17}.

Assuming that, for any $\lambda$ and $n$, the equation ${\zeta =v_{n, \zeta}(\lambda)}$ has a unique solution (no exceptions are known), the solution to Eq.~\eqref{eq:cont} is most easily written in terms of the filling functions $\vartheta_n(\lambda)$ as follows 
\begin{equation}
\label{eq:solution_hydro}
\vartheta_{n ,\zeta}(\lambda) = \vartheta_{n, R }(\lambda)\theta_{\zeta -v_{n, \zeta}(\lambda)}+\vartheta_{n, L } (\lambda)\theta_{v_{n, \zeta}(\lambda)-\zeta}\,.
\end{equation}
Here $\theta_x$ is the Heaviside theta function, which is nonzero and equal to $1$ only if $x>0$, while the ``left" and ``right" filling functions $\vartheta_{n, L}(\lambda)$ and $\vartheta_{n, R}(\lambda)$ are those characterizing the state at infinite distance from the junction on the right and on the left hand side, respectively. In our case, they correspond to thermal states with inverse temperatures $\beta_{L}$ and $\beta_{R}$, and read as
\be
\vartheta^{th}_{n,L/R}(\lambda)= \frac{1}{1+e^{\beta_{L/R} \varepsilon^{th}_{n,L/R}(\lambda)}}\,,
\ee
where $\varepsilon^{th}_{n,L/R}(\lambda)$ is the thermal dressed energy, which satisfies
\begin{align}
\label{eq:therm}
\varepsilon_{n,L/R}^{th}&(\lambda) = e_n(\lambda)+h_{L/R}\nonumber\\
&+\beta^{-1}\left[ \sum_{m=1}^\infty \!T_{nm} \ast \ln\left[ 1 + e^{-\beta\varepsilon_{m,L/R}^{th}}\right]\right]\!(\lambda)\,.
\end{align} 
We stress that the solution \eqref{eq:solution_hydro} is implicit: it depends on $v_{n, \zeta}(\lambda)$, which in turn depends on the state. These equations can be generally solved numerically by simple iterative schemes.  

Ref.~\cite{BCDF16} focused on the XXZ chain for ${|\Delta|<1}$. The derivation proposed is very general and can be applied also to the XXZ chain for ${\Delta>}1$; \eqref{eq:cont} continues to hold also there. 
To completely characterize the states, however, there is a missing ingredient: we need to understand the behavior of the sign $\frak f_\zeta$ 
(\emph{cf}. Sec.~\ref{subsec:charges}). Only once the behavior of $\frak f_\zeta$ is known, the hydrodynamic description becomes complete. This problem will be addressed in the next section. 

\subsubsection{Light Cones}\label{sec:lightcones} %

The structure of the solution \eqref{eq:solution_hydro} allows us to infer some general properties of the profiles of the local observables as a function of the ray $\zeta$. To that aim, let us consider a ray ${\zeta < \min_{\lambda} [v_{n,\zeta}(\lambda)]}$, that is to say $\zeta <v_{n,\zeta}(\lambda)$ for any $\lambda$. From \eqref{eq:solution_hydro} it follows that the state at that ray has no information about the bound states of type $n$ coming from the right hand side. 
Since we assumed that the equation $\zeta=v_{n,\zeta}(\lambda)$ has a unique solution and $\lim_{\zeta \rightarrow -\infty} v_{n,\zeta}(\lambda)$ is finite, we have
\be\label{eq:unique}
\zeta< v_{n,\zeta}(\lambda)\Leftrightarrow\zeta< \bar {\zeta}_n(\lambda)\qquad \forall \lambda
\ee 
where $\bar {\zeta}_n(\lambda)$ is the solution to the equation
\be
 \bar {\zeta}_n(\lambda)= v_{n, \bar {\zeta}_n(\lambda)}(\lambda)\qquad \forall\lambda\, .
\ee
Using \eqref{eq:unique}, one can easily prove  
\be
\zeta < \min_{\lambda} [v_{n,\zeta}(\lambda)]\Leftrightarrow \zeta<\min_\lambda\bar\zeta_{n}(\lambda)
\ee
and 
\be
\label{eq:zeta-}
\min_\lambda\bar\zeta_{n}(\lambda)=\zeta_n^-\, ,
\ee
where $\zeta_n^-$ is the solution to the equation
\be
\zeta_n^-=\min_\lambda [v_{n,\zeta^-_n} (\lambda)]\, .
\ee
We call ${\zeta}^-_n$  the ``negative $n$-th light cone".  The ray ${\zeta}^-_n$ is the one where the first particles of type $n$ coming from the right become visible. Analogously, we define the ``positive $n$-th light cone" ${\zeta}^+_n$ as the solution to the equation
\be
\label{eq:zeta+}
{\zeta}^+_n = \max_{\lambda} [v_{n, {\zeta}^+_n}(\lambda)]\, .
\ee
For $\zeta>{\zeta}^+_n$, there is no bound state of type $n$ coming from the left hand side.

When $\zeta$ is close to $\zeta_n^\pm$,  the profiles of the local observables have a typical square root behavior  $\braket{\boldsymbol{\mathcal O}} \sim \theta_{\mp\zeta\pm \zeta_n^\pm}\sqrt{\mp\zeta\pm \zeta_n^\pm}$. These non-analytic points are  visible in the numerical solutions of \eqref{eq:cont}, see, \emph{e.g.}, Figs~\ref{fig:profiles1} and \ref{fig:correlationprofiles}. This is essentially the same behavior seen in noninteracting models, where, close the light cones, observables display universal properties that belong to the KPZ universality class~\cite{EiRz13,ADSV16}.

Finally, we note that, quite generally, the images of the velocities shrink in the limit of large $n$, and the velocities converge to a constant $\lim_{n \to \infty} v_{n,\zeta}(\lambda) = v_{\infty,\zeta}$ independent of $\lambda$. As a consequence, the following limits exist
\be
\lim_{n\rightarrow\infty}\zeta_n^+=\lim_{n\rightarrow\infty}\zeta_n^-=\zeta_\infty\, .
\ee
In the following, we analyze in detail the behavior of space-time profiles of local observables in correspondence of this ray. In particular, we show that odd operators might exhibit a discontinuous behavior depending on the initial state, signalling the presence of non-ballistic transport. Further details on the fine structure of the profiles near $\zeta_{\infty}$ will be presented elsewhere.

\section{Ballistic and non-ballistic transport for $\Delta\geq1$}\label{sec:main} %

The XXZ model \eqref{eq:Hamiltonian_XXZ} is integrable and, like any other integrable model, is characterized by excitations that propagate ballistically. This allowed us to develop the hydrodynamic theory presented in Section~\ref{sec:hydrodynamiceq} as a kinetic theory of particle excitations moving throughout the system. However, in some cases, symmetries may prohibit ballistic transport of certain quantities, leading to sub-ballistic (such as diffusive) behavior.  This happens in the gapped regime $|\Delta| \geq 1$, where the spin-flip invariance of the root densities \eqref{eq:chargesrho} results in a non-ballistic propagation of the spin degrees of freedom. 
Specifically, there is a region $\cal D$, of size  ${|{\cal D}|\sim t^{\alpha}}$ with $\alpha<1$, where the magnetization experiences finite variations. 
Clearly, the magnetization profile as a function of the ray $\zeta=x/t$ becomes discontinuous at the ray $\bar \zeta$ corresponding to the region $\cal D$. The description of the sublinear scaling region $\cal D$ goes beyond hydrodynamics, and most of the past investigations have been numerical \cite{McCoyDiffusion,Znid11,ProsenDMRG,LowerBoundProsen}. Here we show that the hydrodynamic picture can provide useful information even in such cases. In particular, we point out that a sub-ballistic region generically emerges  by joining states with opposite total magnetization. Moreover, we demonstrate that such sub-ballistic behavior does not correspond always to the NESS region $\zeta=0$, but it can be developed at finite rays $\bar \zeta$.

\subsection{The sign of the odd operators}\label{sec:signodd} %

Let us focus on the case where the two halves of the initial state have magnetizations of opposite signs, $\frak f_L\frak f_R<0$ (\emph{cf}.~\ref{subsec:charges}). For our initial state \eqref{eq:ini_state}, this situation is realized when $h_Lh_R<0$.  Here we show that the profiles of all local operators $\boldsymbol{\mathcal{O}}$ that are odd under spin-flip develop a discontinuity at a given ray $\bar{\zeta}$, as clearly visible in Fig.~\ref{fig:jumps_gapped}. More precisely, we prove that the sign $\frak f_\zeta$ has a single discontinuity at a ray $\bar\zeta$, whose position is fixed by the rapidity distributions $\rho_n(\lambda)$ of the left and right states. 

We start by considering the continuity equation for the magnetization
\be
\zeta \partial_\zeta \bigl(\mathfrak{f}_\zeta\braket{\boldsymbol{s^z}}_{\zeta}^+\bigr)=\partial_\zeta \bigl(\mathfrak{f}_\zeta \braket{\boldsymbol{J}_s}_{\zeta}^+\bigr)\, ,
\label{eq:gencontspin}
\ee
where $\braket{\boldsymbol{s^z}}_{\zeta}^+$ and $\braket{\boldsymbol{J}_s}_{\zeta}^+$ are the expectation values in a state with positive magnetization; they are given by Eqs~\eqref{eq:magn-rhos} and \eqref{eq:Jspin-rhos}. From the continuity equation \eqref{eq:cont} for the root densities it follows 
\be
\zeta \partial_\zeta\braket{\boldsymbol{s^z}}_{\zeta}^+=\partial_\zeta \braket{\boldsymbol{J}_s}_{\zeta}^+\,.
\label{eq:contspin+}
\ee
Using this in \eqref{eq:gencontspin} we find  
\be\label{eq:continuity_s}
(\braket{\boldsymbol{J}_s}_{\zeta}^+ -\zeta \braket{\boldsymbol{s^z}}_{\zeta}^+) \partial_\zeta \mathfrak{f}_\zeta=0\,.
\ee
The solution to this equation is a piecewise constant function of $\zeta$ equal to $\pm 1$ ($\mathfrak{f}_\zeta$ is a sign), which can be written as  
\begin{equation}\label{eq:solution_hydro-sign}
\mathfrak{f}_\zeta= \mathfrak{f}_R\, \theta_{\zeta-v_\zeta^z}+\frak{f}_L\,\theta_{v_\zeta^z-\zeta}\,.
\end{equation}
Here we have used that the equation 
\be\label{eq:spinflipequation}
\zeta=v_\zeta^z\equiv \frac{ \braket{\boldsymbol{J}_s}_{\zeta}^+}{\braket{\boldsymbol{s^z}}_{\zeta}^+}\,,
\ee
has a unique solution. This can be proved by integrating the continuity equation \eqref{eq:contspin+}, which gives 
\be
\zeta-v_\zeta^z=\frac{\int_{\bar\zeta}^\zeta\braket{\boldsymbol{s^z}}_{\zeta}^+}{\braket{\boldsymbol{s^z}}_{\zeta}^+}\, ,
\ee
where we called $\bar\zeta$ a zero of $\zeta-v^z_\zeta$. Since, by definition, ${\braket{\boldsymbol{s^z}}^+\geq 0}$, the right hand side is equal to zero only for $\zeta=\bar\zeta$, that is to say,  the solution to \eqref{eq:spinflipequation} is unique.

The solution $\bar \zeta$ has a nice interpretation in terms of light cones (\emph{cf}. Sec.~\ref{sec:lightcones}). Considering the velocity $v^z_\zeta$ and using the identities \eqref{eq:magn-rhos} and \eqref{eq:Jspin-rhos}, we have 
\be
v^z_\zeta=\lim_{n\rightarrow\infty}\frac{\int\mathrm d \lambda\, v_{n,\zeta}(\lambda)\rho_{n,\zeta}^h(\lambda)}{\int\mathrm d \lambda\,\rho_{n,\zeta}^h(\lambda)}\, .
\ee
As observed in Section~\ref{sec:lightcones}, the images of the velocities shrink in the limit of large $n$ (\emph{cf}. Sec~\ref{sec:lightcones}), thus we find
\be\label{eq:vzvinf}
v_\zeta^z = v_{\infty,\zeta}\, .
\ee
The solution $\bar \zeta$ to \eqref{eq:spinflipequation} is then identified with the accumulation point $\zeta_\infty$ for the light cones.

Eqs \eqref{eq:solution_hydro-sign} and  \eqref{eq:vzvinf}, together with \eqref{eq:solution_hydro}, fully characterize the state in the hydrodynamic limit. Despite the notion of $\mathfrak f$ is outside the TBA description, Eq.~\eqref{eq:vzvinf} suggests that the information about the sign of the odd operators is carried by the heaviest bound state. This result is not surprising if one looks at the behavior of the spin density and related current in the gapless regime ($|\Delta|<1$) for root of unity points $\Delta=\cos(\pi/n)$, with $n$ an integer number. In that case there are $n$ species of excitations and the information about the sign of the magnetization is encoded in the last two species~\cite{DeLuca16}. This can be seen in Fig.~\ref{fig:gaplessVSgap}, in the cases $\Delta=\cos(\pi/3)$ and $\Delta=\cos(\pi/7)$: the spin density and current do not change sign before the particles of the last two species (which have the same velocities) have become visible. In the limit $\Delta \rightarrow 1^-$, $n$ approaches infinity, and the last two species are sent to infinity. If this property does not break down in the gapped regime, the sign of the odd operators should not change before the light cones of the heaviest bound states. Since, for $\Delta>1$, the corresponding velocities approach a constant, $\zeta_\infty$ has to be exactly the ray where the sign changes.

Remarkably, the sign of the front's velocity can give global information about the magnetization profile. For example, if the front moves towards the side with larger magnetization (in modulus), the absolute value of the magnetization can not be monotonous inside the light cone. This can be proved by \emph{reductio ad absurdum}. Let us assume that  the absolute value of the magnetization is smaller on the right hand side, so the front is propagating to the left, \emph{i.e.} it has negative velocity. If $\braket{\boldsymbol{s^z}}_{\zeta}^+$ is monotonous, using the continuity equation \eqref{eq:contspin+}, we have
\be\label{eq:monots}
0\geq |\zeta|\partial_\zeta\braket{\boldsymbol{s^z}}_{\zeta}^+=\mathrm{sgn}\zeta\,  \partial_\zeta\braket{\boldsymbol{J}_s}_{\zeta}^+\, .
\ee
Integrating this equation from $-\infty$ to the accumulation point $\bar\zeta$ gives
\be
\label{eq:absurd}
\braket{\boldsymbol{J}_s}_{{\bar\zeta}}^+\geq 0\, ,
\ee
where we used that the current outside the light cone is zero. The inequality in \eqref{eq:absurd} can not be satisfied because $\braket{{\boldsymbol{J}_s}}_{{\bar\zeta}}^+$ has the sign of $\bar\zeta=v^z_{\bar\zeta}$, which was negative by assumption (\emph{cf}. Eq.~\eqref{eq:spinflipequation}). 

\section{Results}\label{sec:results}  %

In this section we elaborate on our predictions for the profiles of local observables as a function of the ray $\zeta$ and show a comparison with time-dependent density matrix renormalization group (tDMRG) simulations. 
The predictions are obtained by taking the expectation value of local observables in the state $\boldsymbol \rho_s(\zeta)$, which we represent microcanonically by $\ket{\rho_\zeta,\frak f_\zeta}$, where $\rho_\zeta$ and $\frak f_\zeta$ are computed by first solving \eqref{eq:cont} and then \eqref{eq:solution_hydro-sign}. 

The tDMRG simulations are obtained for finite lattices of $L$ sites, with $L \in [80,120]$,  imposing open boundary conditions. 
In order to initialize the system in the state (\ref{eq:ini_state}), we proceed as follows:

{\it (i)} We prepare each half-chain in the mixed product state $\prod_{j} e ^{(\beta h)_{L/R} \boldsymbol s^{z}_{j} }$. In terms of {\it locally purified} matrix product states (MPS), such a state only needs a two-dimensional {\it ancilla}  and an auxiliary bond dimension $\chi = 1$.
 {\it (ii)} We implement imaginary time evolution using second-order 
Trotter decomposition of the operator $\propto e^{-\beta_{L/R} {\boldsymbol H}}$,
with imaginary time-step $d\beta = 10^{-3}$.
{\it (iii)}
We evolve both the initial left and right mixed product states 
up to the desired temperatures.

After joining together the two open chains, the system is 
unitarily evolved using second-order Trotter decomposition of the evolution operator,
with time-step $dt = 10^{-2}$. During the time evolution,
the bond dimension of the MPS is dynamically updated,
up to a maximum value $\chi_{\rm max} = 200$. For this reason,
the maximum time we can reach keeping the accumulated error reasonably small
is $t_{\rm max} \simeq 20$. 

Exploiting the structure of the matrix product state, we can easily measure 
any local observable, including charge densities, currents  
and, more in general, correlation functions. 

\subsection{Homogeneous magnetization signs: light cones}%
\label{sec:homo_sign}

\begin{figure}[h]
\includegraphics[width=0.45\textwidth]{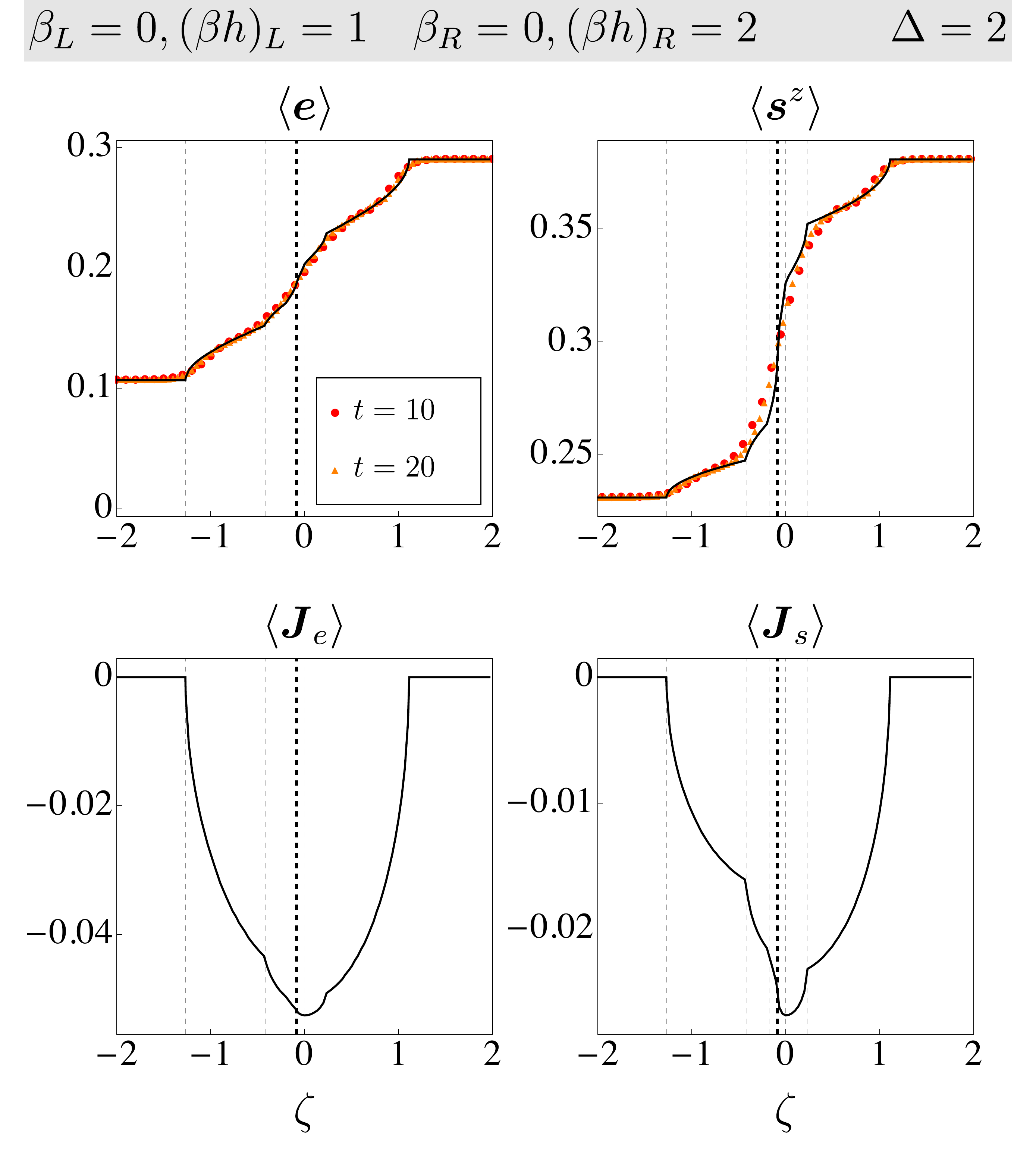}\vspace{-0.25cm}
\caption{Space-time profiles of densities and currents of spin and energy.  Solid black lines display the theoretical predictions, while points correspond to the exact time evolution computed by tDRMG simulations up to times $t=20$. Vertical dashed lines represent positive and negative light cones of the different $n$-quasiparticle bound states, see Sec.~\ref{sec:lightcones}. The corresponding rays $\zeta={\zeta}^{\pm}_n$, with $n=1,2,3$, are displayed as gray dashed lines, while the black dashed line corresponds to the largest string $\zeta={\zeta}_\infty$ .  }
\label{fig:profiles1}
\end{figure}

Let us start by considering the case where the sign of the magnetization is homogeneous in the initial state and $\frak f_\zeta$ is constant throughout the light cone. In this setting the qualitative behavior of the space-time profiles does not differ much from the one in the gapless regime.

In Fig.~\ref{fig:profiles1} we report the space-time profiles of local observables after the sudden junction of two infinite-temperature states with different (but positive) magnetizations. One immediately sees that the profiles are not smooth, presenting a number of cusps. These are the non-analytic points discussed in Sec.~\ref{sec:lightcones}, and their positions $\{{\zeta}^{\pm}_n\}$ can be predicted by solving Eqs~\eqref{eq:zeta-} and \eqref{eq:zeta+}. As discussed in Sec.~\ref{sec:lightcones}, these points have a natural interpretation in terms of moving quasiparticles: $\zeta^{+}_n$ and $\zeta^-_n$ correspond to the rays where the quasiparticles of species $n$, coming respectively from the left and from the right, become visible.

The first light cone is where the profiles begin to deviate from a constant function.
This ray corresponds to the velocity of the fastest particles (the unbound magnons in our case). Note that, since the system is interacting, the maximal velocities on the two sides are generically different from one another. This is the case for the data reported in Fig.~\ref{fig:profiles1}. 
As the dispersion law of quasi-particles is smooth, the profiles are expected to remain smooth between two consecutive light cones.

Cusps are also present in the gapless regime studied in \cite{BCDF16}; depending on the initial state, they can be more or less  marked. 

\begin{figure}[h]
\includegraphics[width=0.45\textwidth]{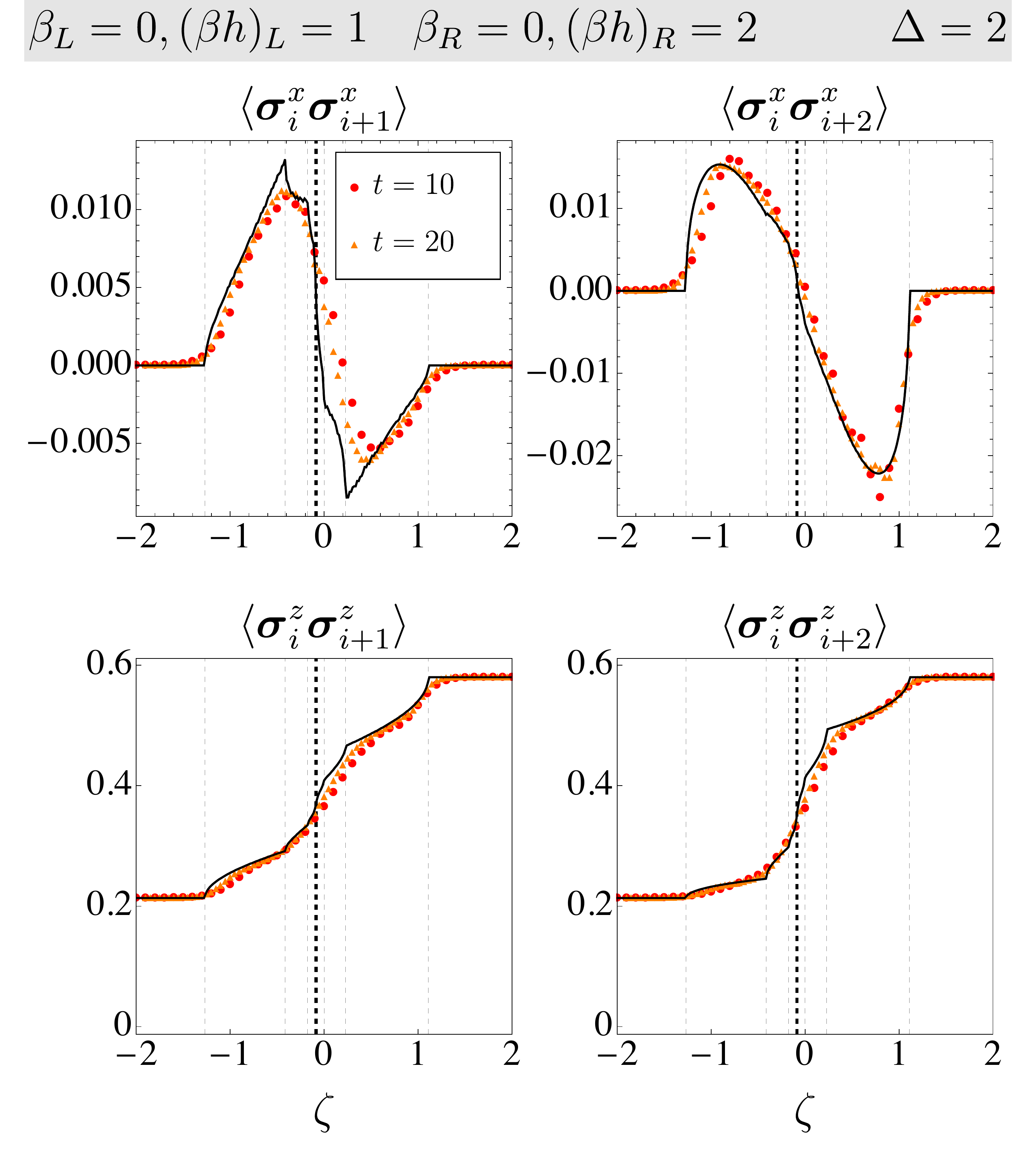}\vspace{-0.25cm}
\caption{ Space-time profiles of local correlators, same notations as in Fig.  \ref{fig:profiles1}. Note that the absolute value of correlators along the $x$-direction is two orders of magnitude smaller than that along the $z$-direction. In the former case the visible small ripples on the theoretical curves are numerical artifacts.}
\label{fig:correlationprofiles}
\end{figure}

As the rapidity distributions $\rho_n(\lambda)$ completely characterize the state, the solution to the hydrodynamic equation \eqref{eq:cont} allows us to investigate further light-cone properties, going beyond the analysis of conserved charges and currents. To that aim, we use some recently developed formulae~\cite{MePo14} for the expectation values of local observables in generic eigenstates of the gapped XXZ Hamiltonian. 
In particular, we have computed nearest and next-to-nearest neighbor correlations inside the light-cone. Our results are reported in Fig.~\ref{fig:correlationprofiles}. Once again, cusps are clearly visible. We also observe an interesting, non-monotonic behavior of transverse correlators. Fig.~\ref{fig:correlationprofiles} also displays data from tDMRG simulations, which are found to be in very good agreement with our predictions, further corroborating the validity of our results.

\begin{figure}[t]
	\includegraphics[scale=0.37]{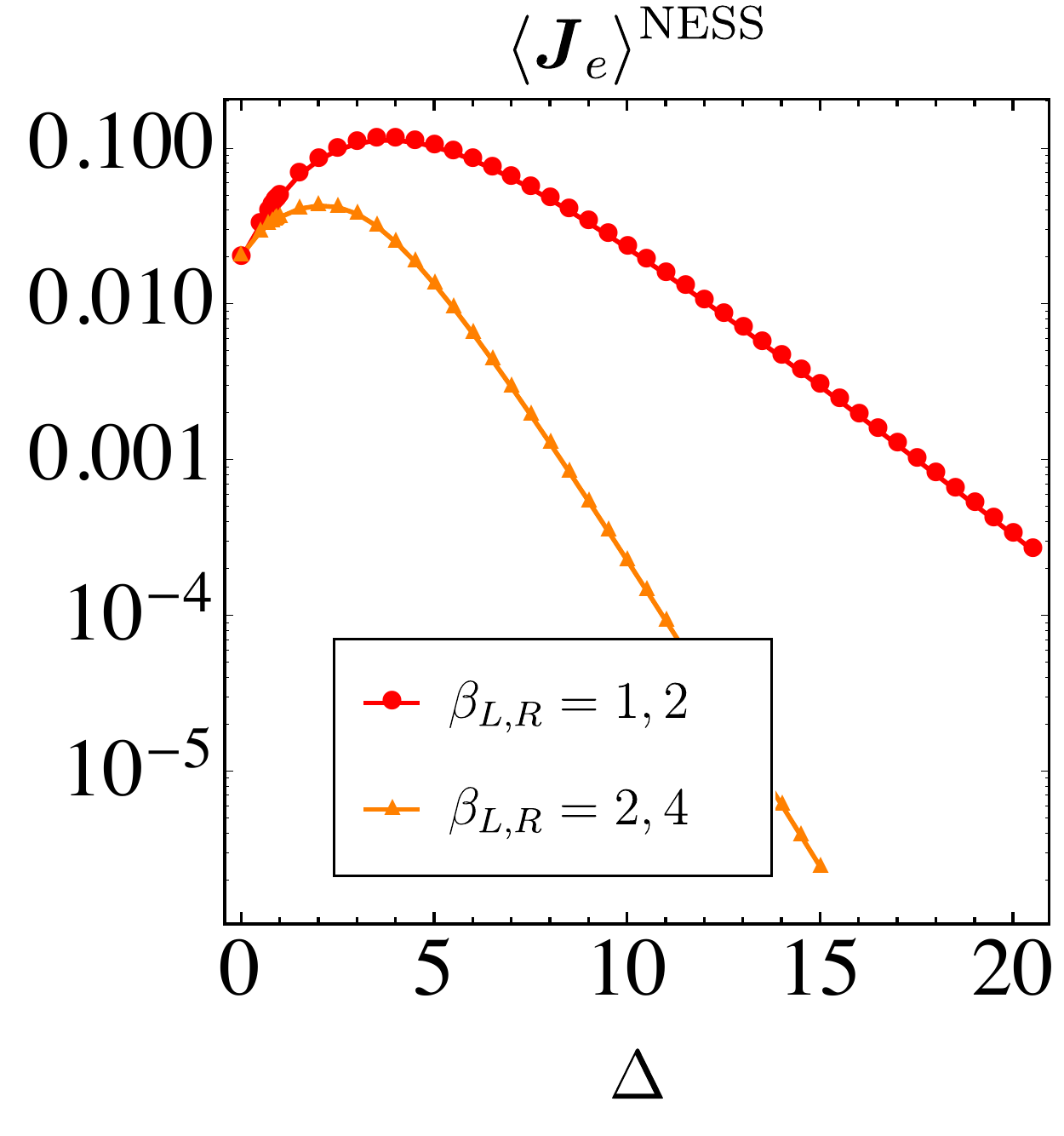}\vspace{-0.25cm}
	\caption{ NESS energy current $\braket{\boldsymbol{J}_e}_{\zeta=0}$ as a function of the anisotropy $\Delta$. The initial state is prepared by joining together two semi-infinite chains with vanishing magnetic field and different temperatures.}
\label{fig:Ness_current}
\end{figure}

Finally, before turning to the next section, we provide a dedicated analysis of the celebrated NESS energy current, corresponding to $\braket{\boldsymbol{J}_e}_{\zeta=0}$. Fig.~\ref{fig:Ness_current} shows its value as a function of the anisotropy $\Delta$, in the case where the two semi-infinite chains are initially prepared at different temperatures and with vanishing magnetic fields. The energy current has a non-monotonic behavior in $\Delta$, reaching a peak when $\Delta \sim \min \bigl( \beta_{R}^{-1},\beta^{-1}_L \bigr)$. Furthermore, for the chosen values of the initial  parameters, the maximum is reached for $\Delta>1$. The current is always seen to vanish exponentially for $\Delta\to\infty$, as one can clearly see from the logarithmic plot in Fig.~\ref{fig:Ness_current}. As a function of the temperatures, it approximately behaves as 
$\sim \exp\bigl[- \Delta\min \bigl( \beta_{R},\beta_L \bigr)/2 \bigr]$.

\subsection{Heterogeneous magnetization signs: spin-jumps}  %

We now turn to presenting our results for the case where the semi-infinite spin chains are initially prepared in equilibrium states with different magnetization signs $\mathfrak{f}_R  = - \mathfrak{f}_L \equiv - \mathfrak{f}$. 

\begin{figure}[h]
\includegraphics[width=0.45\textwidth]{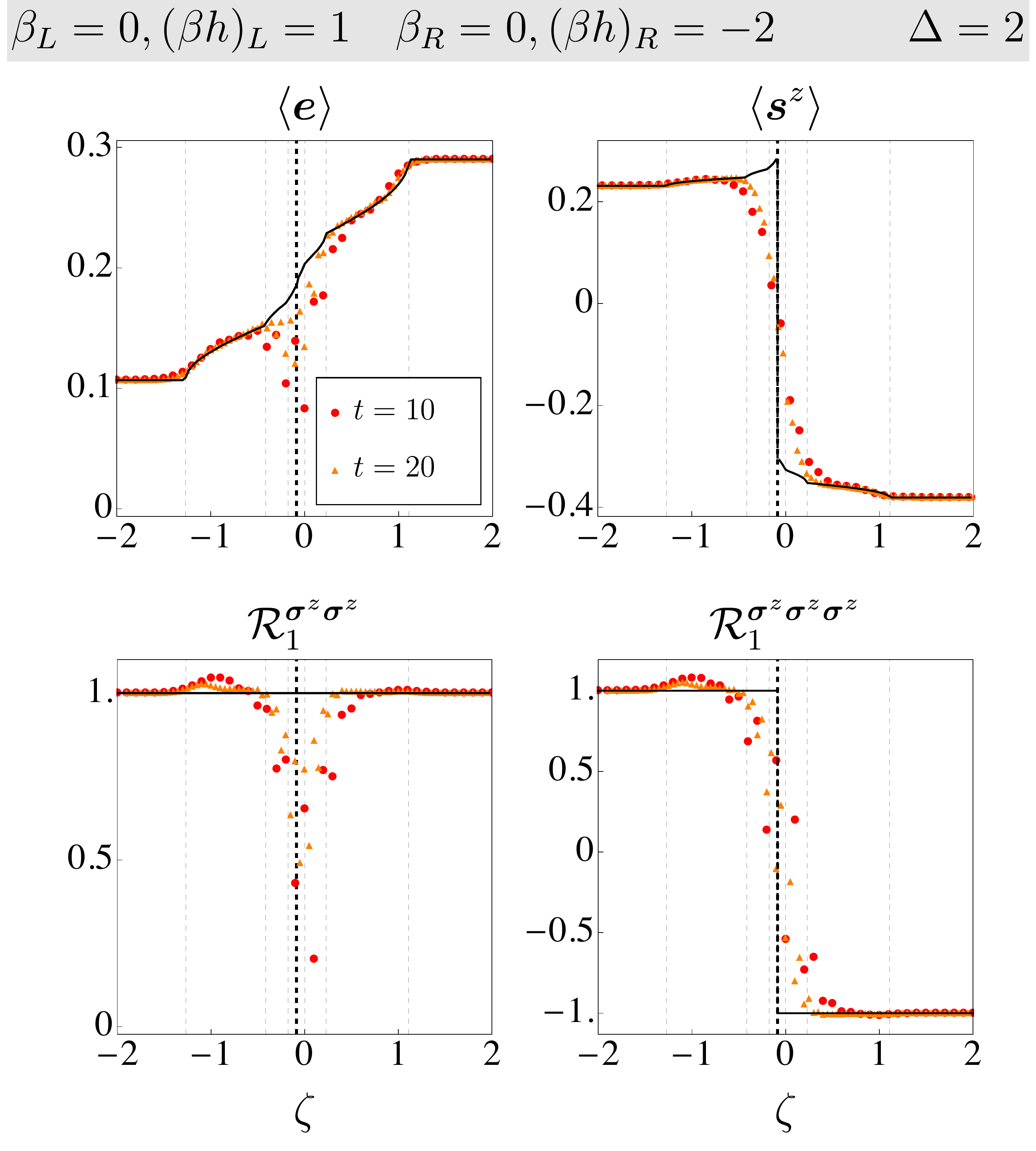}\vspace{-0.25cm}
\caption{ Space-time profiles of energy and magnetization (top) and spin-spin correlation functions (bottom), same notations as in Fig.  \ref{fig:profiles1}. The function $\mathcal{R}_1^{\boldsymbol\sigma^z \boldsymbol\sigma^z}$ is computed as the ratio between two different profiles: the first is the profile of the correlator $\langle\boldsymbol\sigma^z_i\boldsymbol\sigma^z_{i+1}\rangle$ obtained by joining thermal states with $\mu_L =1$, $\beta_L=0$ and $\mu_R = -2$, $\beta_R=0$; the second is the profile for $\mu_L =1$, $\beta_L=0$ and $\mu_R = +2$, $\beta_R=0$. The plot for $\mathcal{R}_1^{\boldsymbol\sigma^z \boldsymbol\sigma^z\boldsymbol\sigma^z}$ is obtained analogously from the correlator $\langle \boldsymbol{\sigma}^z_i \boldsymbol{\sigma}^z_{i+1}  \boldsymbol{\sigma}^z_{i+2} \rangle$. Note that the odd operators show a genuine discontinuity at $ \zeta_{\infty} \sim -0.086$ (black vertical dashed line).}
\label{fig:jumps_gapped}
\end{figure}

In light of the discussion in Sec.~\ref{sec:signodd}, we expect the observables that are odd under spin-flip to abruptly flip their sign at $\zeta=\zeta_\infty$. This is nicely displayed in Fig.~\ref{fig:jumps_gapped}, where we reported our theoretical predictions and numerical data from tDMRG simulations.  

In order to test our predictions of the jumps against numerics, we proceed as follows. We fix a local observable $\boldsymbol{\cal O}_i$ and compute its profiles starting from two different initial states $\boldsymbol\rho^{(1)}_0$ and $\boldsymbol\rho^{(2)}_0$. These states are chosen to differ only in the sign of the magnetic field on the right. For the first state we have $\frak f_L=\frak f$ and $\frak f_R=\frak f$, while, for the second one, $\frak f_L=\frak f$ and $\frak f_R=-\frak f$. We then define the ratio 
\begin{equation}
\mathcal{R}_{\mathfrak{f},\zeta}^{\boldsymbol{\mathcal{O}}}(t) \equiv \frac{\tr{}{\boldsymbol{\mathcal{O}}_{\zeta t}(t)\boldsymbol \rho^{(2)}_0}}{\tr{}{\boldsymbol{\mathcal{O}}_{\zeta t}(t)\boldsymbol \rho^{(1)}_0}}\,.
\end{equation}
This ratio is such that 
\be
\lim_{t\rightarrow \infty} \mathcal{R}_{\mathfrak{f},\zeta}^{\boldsymbol{\mathcal{O}}}(t) =\begin{cases}
\text{sgn}(\zeta_\infty-\zeta)& \Pi \boldsymbol{\cal O}_i\Pi=-\boldsymbol{\cal O}_i\\
1&\Pi \boldsymbol{\cal O}_i\Pi=\boldsymbol{\cal O}_i\, .
\label{eq:sing_prediction}
\end{cases}
\ee
We remark that the analytic calculation of the profile of $\boldsymbol{\cal O}_i$ is not required to test this prediction; this allows us to consider also observables for which we are not able to calculate the profile. For example, in Fig.~\ref{fig:jumps_gapped} we also report $\mathcal{R}_{\mathfrak{f},\zeta}^{\boldsymbol{\mathcal{O}}}(t)$ for ${\boldsymbol{\cal O}_i=\boldsymbol{\sigma}^z_i \boldsymbol{\sigma}^z_{i+1}  \boldsymbol{\sigma}^z_{i+2}}$. We see from the figure that we are able to successfully test \eqref{eq:sing_prediction} against tDMRG data, even though no formula involving the root densities is currently available for computing the expectation value of this operator.

In all the cases considered, the tDMRG simulations are compatible with our predictions, but the corrections are not always small. 
In particular, a slow, sub-ballistic behavior is expected at the discontinuity of the profiles, which contributes to the presence of large finite-time effects. As a result, the tDMRG simulations can not reach sufficiently long times to observe an actual discontinuous behavior. We ascribe the differences between our predictions and the tDMRG data to such numerical problems; our analysis of how the tDMRG data approach their asymptotic values supports that conclusion.

\begin{figure}[t]
\includegraphics[width=0.45\textwidth]{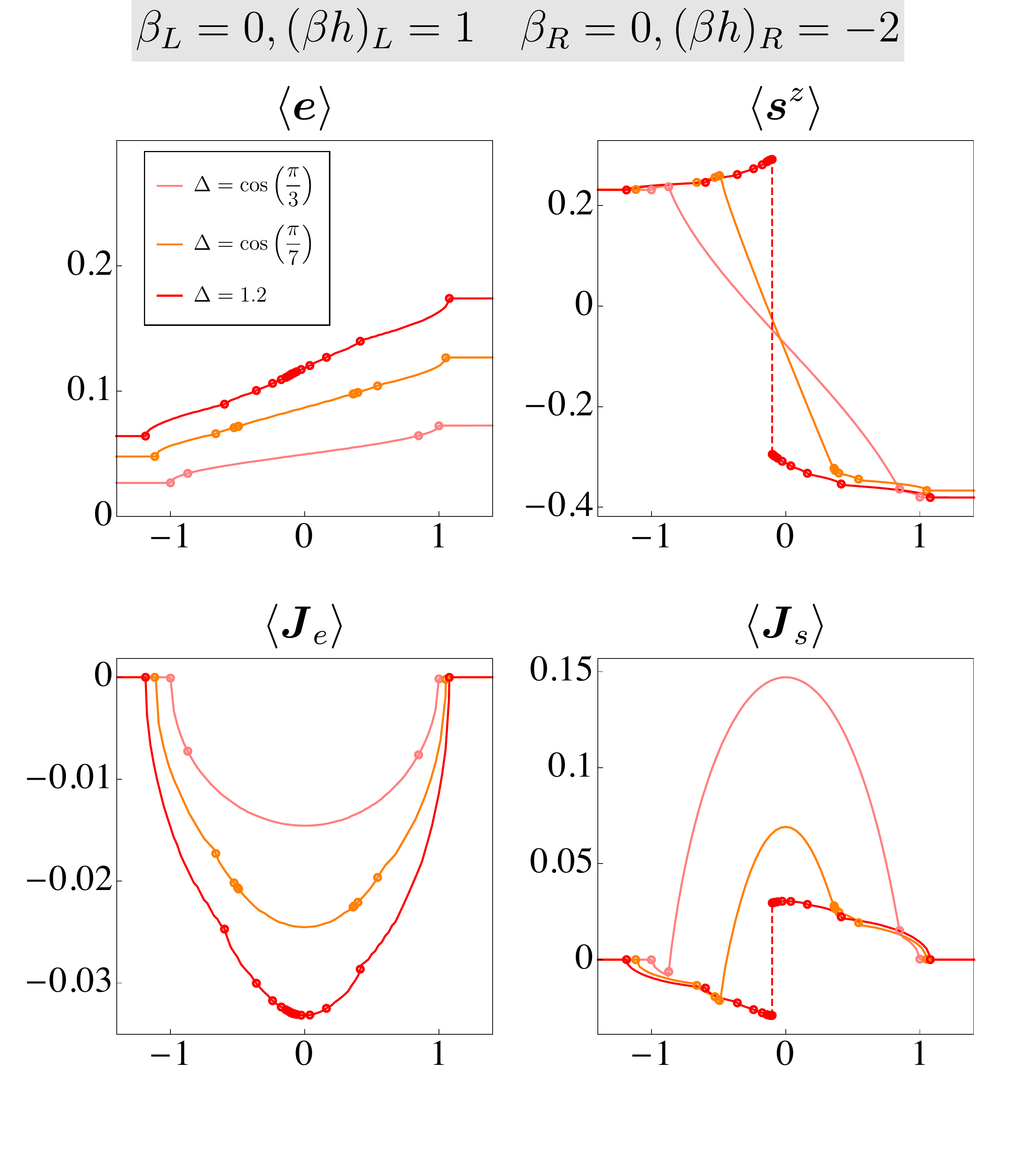}\vspace{-0.25cm}
\caption{Space-time profiles of densities and currents of spin and energy. Different plots correspond to different values of $\Delta$, in the gapless regime $\Delta=\cos \pi/\ell$, with $\ell=3,7$ and in the gapped regime $\Delta=1.2$. The small circles on top of the profiles indicate the positions of the light cones $\zeta_n^{\pm}$ for each values of $\Delta$. Note that the number of light cones in the gapless regime is finite as the number of species is also finite. In the gapped regime instead there is an infinite number of light cones converging to the ray $\zeta_\infty$, where the magnetization density and the spin current change sign. }
\label{fig:gaplessVSgap}
\end{figure}

The abrupt jumps in the profiles of odd observables displayed in Fig.~\ref{fig:jumps_gapped} find no correspondence in the gapless regime. It is then important to understand how such discontinuities arise as the value of the anisotropy is continuously varied from $\Delta<1$ to $\Delta>1$. Some data are reported in Fig.~\ref{fig:gaplessVSgap}. We see that, while the profiles remain always continuous for $\Delta<1$, they become increasingly sharp as $\Delta$ is increased, finally developing a discontinuity at $\Delta=1$.

\subsection{Zero to finite magnetization: sharp front}%

In this section we finally consider the situation where one of the two semi-infinite chains (say, the left one) is initially prepared in a thermal state with vanishing magnetic field, while the other (the right one) has a non-zero magnetic field. This is a limiting case of the ones presented in the previous subsections. For this quench protocol, the long-time magnetization profiles have definite sign as a function of the ray $\zeta$. Accordingly, the profiles of all local observables are simply obtained from the solution to the hydrodynamic equation \eqref{eq:cont}, in analogy to the situation discussed in Sec.~\ref{sec:homo_sign}. In this case, however, the solution displays some interesting properties which are worth discussing in a detailed fashion. 

\begin{figure}[h]
\includegraphics[width=0.45\textwidth]{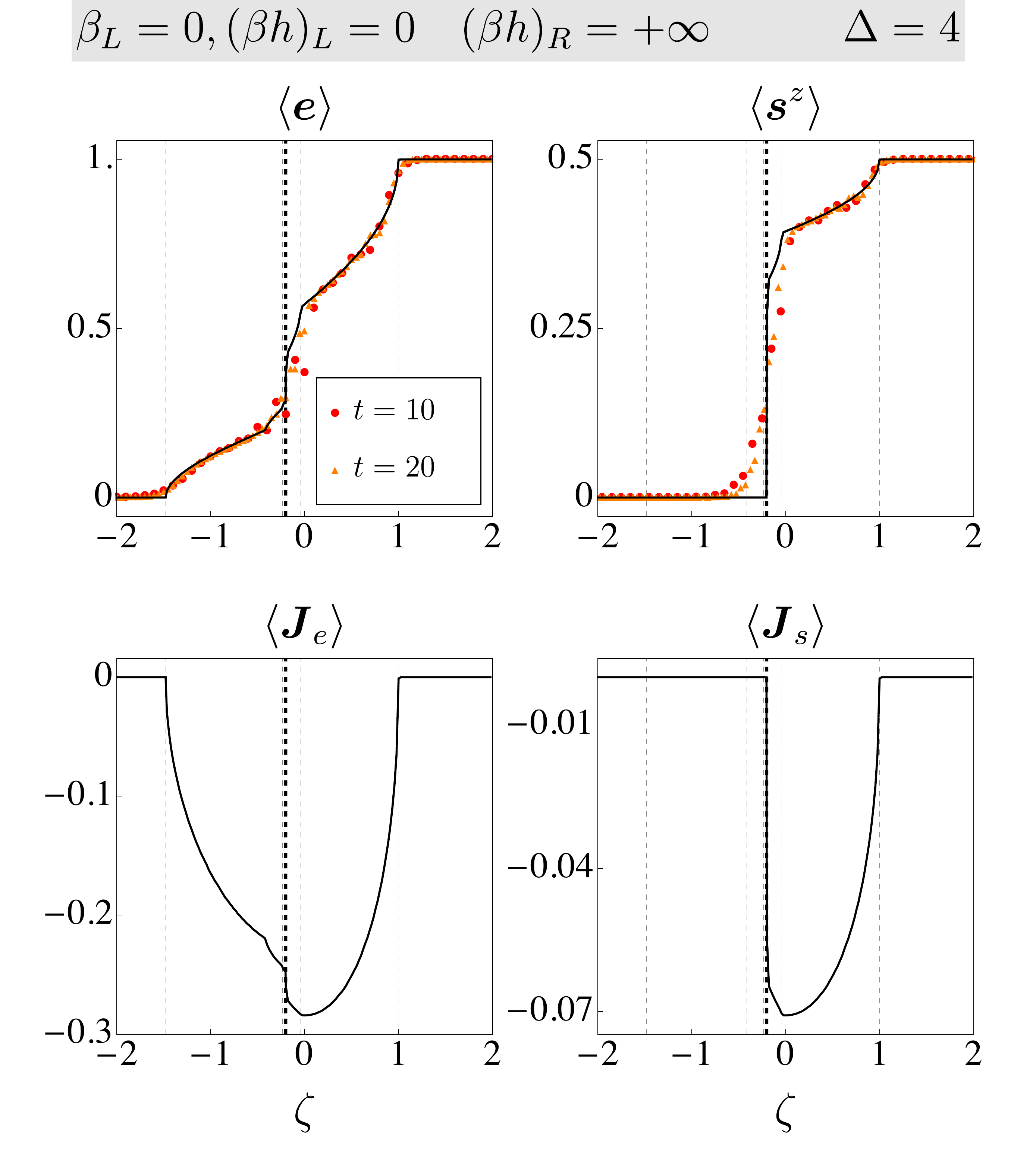}\vspace{-0.25cm}
\caption{Space-time profiles of densities and currents of spin and energy, same notations as in Fig.~\ref{fig:profiles1}. Remarkably, we see that the leftmost light cones of the magnetization and of the energy profiles do not coincide. This is due to the special properties of the initial state, as explained in detail in the main text.}
\label{fig:profiles00}
\end{figure}

\begin{figure}[h]
\includegraphics[width=0.45\textwidth]{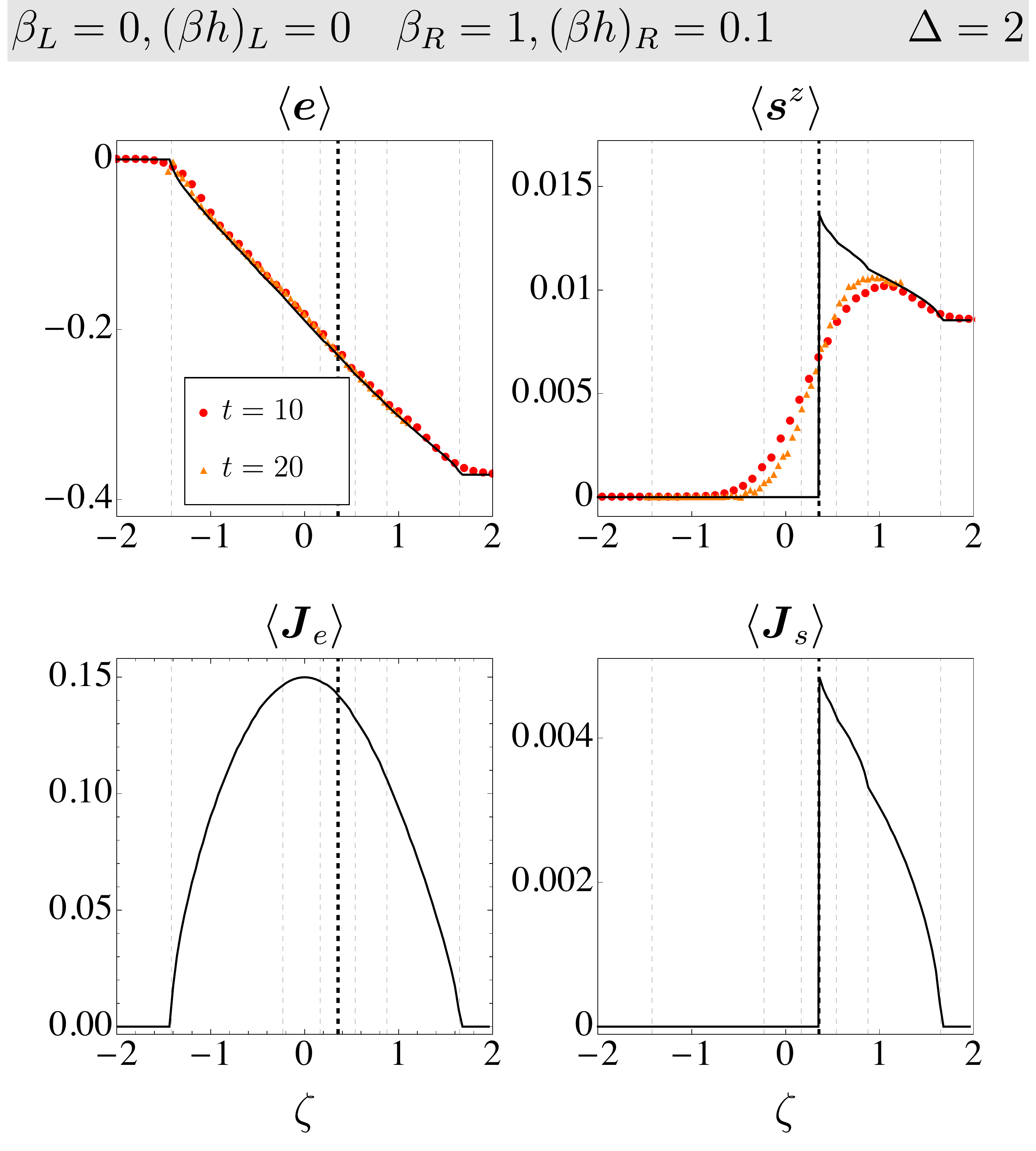}\vspace{-0.25cm}
\caption{Space-time profiles of densities and currents of spin and energy, same notations as in Fig.~\ref{fig:profiles1}. The magnetization exhibits a non-monotonic behavior, which is naturally interpreted as a thermoelectric effect.}
\label{fig:profiles01}
\end{figure}

The first example is a problem of release into the vacuum. The right part of the system is prepared in the state with all spins up (the vacuum),  while the left part is in an infinite temperature state with vanishing magnetic field. The numerical solution to the hydrodynamic equations \eqref{eq:cont} is displayed in Fig.~\ref{fig:profiles00}. We clearly see that the leftmost light cones of the magnetization and energy profiles do not coincide. This remarkable property can be seen as a corollary of our theory on the sign of the odd operators.

In order to show this, we consider the two situations where tiny magnetic fields, respectively positive ($h_L=h_{\epsilon}$) and negative (${h_L=-h_{\epsilon}}$), are initially turned on in the left semi-infinite chain. On the left hand side of 
the first light cone $\zeta_1^-$ the magnetization will be non-vanishing, $\braket{\boldsymbol{s^z}}^{\pm}_{-\infty}=\pm\epsilon$. By integrating the continuity equation \eqref{eq:monots} for the magnetization from $\zeta_1^-$ to $\zeta_1^+$, we find
\be
\pm \int_{\zeta_1^-}^{\zeta_\infty}\mathrm d z\braket{\boldsymbol{s^z}}^+_{z}+\int^{\zeta_1^+}_{\zeta_\infty}\mathrm d z\braket{\boldsymbol{s^z}}^+_{z}=\zeta_1^+\braket{\boldsymbol{s^z}}_{\zeta_1^+}^+\mp \zeta_1^-\epsilon\, ,
\ee
where we used \eqref{eq:solution_hydro-sign} and that the spin current is zero outside the light cone. 
Taking the difference between the two cases gives
\be\label{eq:intzero}
\lim_{\epsilon\rightarrow 0}\int_{\zeta_1^-}^{\zeta_\infty}\mathrm d z\braket{\boldsymbol{s^z}}^+_{z}=0\, .
\ee 
Since $\braket{\boldsymbol{s^z}}^+_{\zeta}$ is nonnegative, \eqref{eq:intzero} implies 
\be
\lim_{h_{\epsilon}\rightarrow 0}\braket{\boldsymbol{s^z}}^+_{\zeta}=\lim_{\epsilon\rightarrow 0}\braket{\boldsymbol{s^z}}^+_{\zeta}=0\qquad\forall \zeta<\zeta_\infty\, .
\ee
It is now reasonable to assume that the magnetization profile for $h_{L} = 0$ can be obtained as the limit $h_{\epsilon}\to 0$ of the profile where the left magnetic field is positive but small. In fact, this is actually implicit in the numerical solution to the hydrodynamic equation \eqref{eq:cont}. This simple argument shows that the magnetization profile for $h_{L}=0$ is vanishing for all the rays $\zeta$ smaller than $\zeta_{\infty}$. 

This is a general property, and is observed every time the initial state has vanishing magnetization on one of its two halves. For example, this is also the case displayed in Fig.~\ref{fig:profiles01}, where the right magnetic field is finite. 

In Fig.~\ref{fig:profiles01}, the magnetization profile exhibits an interesting non-monotonic behavior. The latter is naturally interpreted as a thermoelectric effect and is observed also when the initial halves of the chain have the same non-vanishing magnetization but one part is much colder than the other. 

Finally, we point out that the magnetization profile in Fig.~\ref{fig:profiles01} seems to develop a discontinuity at the accumulation point $\zeta_{\infty}$. Our numerical analysis of the profiles with increasing right magnetic fields seems to suggest that the front could in fact be continuous, albeit extremely sharp. Near the accumulation point $\zeta_{\infty}$, the profile varies very quickly over a region that approaches zero in the limit where the right magnetic field is sent to zero. This could be at the basis of the apparent discontinuity displayed in~Fig.~\ref{fig:profiles01}.

\section{Conclusions}\label{sec:conclusions}%

We have considered the time evolution of local observables after the junction of two semi-infinite XXZ chains initially prepared in thermal equilibrium with different temperatures and magnetic fields. We focused on the gapped regime $|\Delta|>1$, where interactions are larger in the direction of the anisotropy. Our analysis is complementary to the one of \cite{BCDF16}, where the gapless case $|\Delta|<1$ was studied.\\

By means of tDMRG simulations and hydrodynamic equations, we analyzed in detail the emergent space-time profiles of conserved charges, currents, and local correlations. 
We showed that the particle content of the model can be inferred from the profiles, which display two cusps  for each species of excitations.  
Moreover, we showed that sub-ballistic behavior emerges considering observables that are odd under spin flip in the case where the magnetizations in the two semi-infinite chains have opposite signs. Specifically, the long-time profile of odd observables exhibits abrupt jumps, which signal the presence of non-ballistic transport. The location of the jumps, $\zeta_\infty$,  can not be predicted directly from the  hydrodynamic equations satisfied by the root densities. Our main result is to derive an equation that describes the ray $\zeta_\infty$, completing the characterization of the long-time steady profiles of all the local observables. It would be interesting to analyze in greater detail the sub-ballistic behavior around $\zeta_\infty$, as done in \cite{ProsenDMRG} for $\zeta_\infty=0$. This is a challenging numerical problem, which we leave to future investigations.\\

The form of the hydrodynamic equations employed in our work is extremely general and is expected to hold in all Bethe ansatz integrable models or in models where stable particle excitations can be constructed \cite{SMatrixLaurens}. In particular, we expect discontinuities in the space-time profiles of odd operators in every model with root densities invariant under some discrete symmetry. This is, for instance, the case of the Hubbard model \cite{Hubbard_book}, where the root densities are in one-to-one correspondence with spin-flip and charge-flip invariant commuting fused transfer matrices~\cite{IlDe17,Cavaglia15}.\\

Finally, we note that inhomogeneous quantum quenches prove to be an excellent setting where to study the particle content of the model: the space-time profiles of local observables can be used as effective ``spectroscopes'' of collective excitations, as the appearance of singularities in the profiles of the local observables is connected with the presence of more species of quasiparticles. This method is complementary to others suggested for homogeneous quenches, based, for instance, on the computation of local correlations \cite{gristev2007} or entanglement entropy and mutual information \cite{AlCa16, MBPC17}. \\

\begin{acknowledgments} %
We are grateful to J.-S. Caux for drawing our attention to the effects of the bound states in the profiles of local observables. We thank Vincenzo Alba, Benjamin Doyon, Enej Ilievski, and Herbert Spohn for stimulating discussions. \\

B.B. acknowledges the financial support by the ERC under Starting Grant 279391 EDEQS.
J.D.N. and M.F.  acknowledge support by LabEx ENS-ICFP:ANR-10-LABX-0010/ANR-10-IDEX-0001-02 PSL*. M.C. acknowledges support by the Marie Sklodowska-Curie Grant No. 701221 NET4IQ.\\
\end{acknowledgments}%
\paragraph*{Author contributions.}
L. Piroli, J. De Nardis, B. Bertini, and M. Fagotti  contributed equally to the development of the theory; M. Collura took care of the tDMRG simulations. 
In addition, L. Piroli and  J. De Nardis  took charge of the numerical solutions to the hydrodynamic equations; B. Bertini and M. Fagotti played a major role in the organization of the work and in the refinement of the theory.

\end{document}